\documentclass[useAMS, usenatbib]{mnras}

\usepackage[T1]{fontenc}

\DeclareRobustCommand{\VAN}[3]{#2}
\let\VANthebibliography\thebibliography
\def\thebibliography{\DeclareRobustCommand{\VAN}[3]{##3}\VANthebibliography}

\usepackage{graphicx}	% Including figure files
\usepackage{amsmath}	% Advanced maths commands
\usepackage{amssymb}	% Extra maths symbols
\usepackage[export]{adjustbox}
\usepackage{subcaption}
\usepackage{epsfig}
\usepackage{newtxtext,newtxmath}
\title[The extreme redback MSP]{Quantifying irradiation in spider pulsars: the extreme case of PSR~J1622-0315}

\author[Marco Turchetta]{
Marco Turchetta,$^{1}$\thanks{E-mail: marcoturchetta@ntnu.no} Manuel Linares$^{1,2}$, Karri Koljonen$^{1}$ and Bidisha Sen$^{1}$
\\
$^{1}$ Department of Physics, Norwegian University of Science and Technology, NO-7491 Trondheim, Norway\\
$^2$ Departament de F{\'i}sica, EEBE, Universitat Polit{\`e}cnica de Catalunya, Av. Eduard Maristany 16, E-08019 Barcelona, Spain.\\
}

\date{Accepted XXX. Received YYY; in original form ZZZ}

\pubyear{2022}

\begin{document}
\label{firstpage}
\pagerange{\pageref{firstpage}--\pageref{lastpage}}
\maketitle

\begin{abstract}
We present the first multi-band optical light curves of PSR~J1622-0315, among the most compact known redback binary millisecond pulsars, with an orbital period $P_\mathrm{orb}=3.9 \, \text{h}$. We find a flux modulation with two maxima per orbital cycle and a peak-to-peak amplitude $\simeq0.3 \, \text{mag}$, which we attribute to the ellipsoidal shape of the tidally distorted companion star. The optical colours imply a late-F to early-G spectral type companion and do not show any detectable temperature changes along the orbit. This suggests that the irradiation of the star's inner face by the pulsar wind is unexpectedly missing despite its short orbital period.
%We compare this case with the redback PSR~J2215+5135, having a similar orbital period $P_{orb}=4.1 \, \text{h}$ and showing on the contrary strong irradiation, with much deeper optical eclipses of $\approx 1 \, \text{mag}$. These results
%and the observative lack of non-irradiated black widows in contrast to what we see for redbacks ($\sim 50\%/50\%$ irradiated/non-irradiated)
To interpret these results, we introduce a new parameter $f_\mathrm{sd}$, defined as the ratio between the pulsar wind flux intercepted by the companion star and the companion intrinsic flux. This flux ratio $f_\mathrm{sd}$, which depends on the spin-down luminosity of the pulsar, the base temperature of the companion and the orbital period, can be used to quantify the effect of the pulsar wind on the companion star and turns out to be the most important factor in determining whether the companion is irradiated or not.
%In this picture the companion intrinsic temperature, which splits redbacks and black widows in the two ranges $T_\mathrm{b} \simeq$4000--6000~K and $T_\mathrm{b} \simeq$1000--3000~K respectively, turns out to be the most important factor in determining an irradiated or non-irradiated system.
We find that the transition between these two regimes occurs at $f_\mathrm{sd} \simeq 2$--$4$ and that the value for PSR~J1622-0315 is $f_\mathrm{sd}=0.7$, placing it firmly in the non-irradiated regime.
\end{abstract}

\begin{keywords}
 techniques: photometric -- binaries: close -- stars: neutron -- pulsars: general
\end{keywords}

%%%%%%%%%%%%%%%%%%%%%%%%%%%%%%%%%%%%%%%%%%%%%%%%%%

%%%%%%%%%%%%%%%%% BODY OF PAPER %%%%%%%%%%%%%%%%%%

\section{Introduction}
\label{sec:1}
Compact binary millisecond pulsars (CBMSPs) are fast rotating neutron stars, with spin periods down to $P_\mathrm{spin} \sim \text{ms}$, moving in tight orbits ($P_\mathrm{orb}\lesssim1 \, \text{d}$) with lighter companion stars. Due to their short orbital separations, $a \sim \text{R}_{\sun}$, the relativistic wind of high-energy particles emitted by the pulsar can strongly heat and progressively ablate matter from the companion star, fully destroying them in few cases \protect\citep{van1988fate}. This star-destroying behaviour has led to these pulsar systems being nicknamed as \textit{black widows} (BWs), with companion masses $M_{2}\sim 0.01 \, \text{M}_{\sun}$, and \textit{redbacks} (RBs), with $M_{2}\sim 0.3$-$0.7 \, \text{M}_{\sun}$ \citep{d2001eclipsing, roberts2012surrounded}. These \textit{spiders} provide a unique opportunity to study the intra-binary shock that is thought to form between the pulsar wind and the companion wind \citep{wadiasingh2018pressure}. Moreover, they represent ideal sites to find super-massive neutron stars \citep[e.g.][]{kaplan2013metal, linares2018peering, 2020mbhe.confE..23L}, as they have been accreting mass from the companion star over $\sim10^{9}$ years.
%Also performing “blind searches” for pulsations in the gamma-ray band turns out to be very challenging, owing to the huge computing power required for searching the signal testing a wide range of orbital parameters \citep{pletsch2011discovery}.

Unfortunately, detecting CBMSPs in radio pulsar surveys is quite challenging, due to eclipses of the pulsations occurring at a large fraction of the orbit \citep{d2001eclipsing}. Therefore, in recent years, many spiders have been discovered and studied observing their optical counterparts, which typically show a peculiar large-amplitude flux modulation due to the companion orbiting around the pulsar \citep{breton2013discovery, linares2017millisecond, strader2019optical, swihart2021discovery}. Such periodic emission variability is induced by the combination of the effect of the non-spherical shape of the companion, tidally distorted by the neutron star's gravitational field, and the strong irradiation of the companion's inner face by the pulsar wind.
%Studying the optical counterparts of spiders turns out to be very useful also to constrain important parameters of the system, specially the neutron star mass, by modeling the light curves and eventually combine these with spectroscopic radial velocity measurements, to infer the parameters more accurately \citep{shahbaz2017properties, strader2019optical}.

Here, we present the first multi-band optical light curves of PSR~J1622-0315 (henceforth called J1622), one of the most compact known RBs with $P_\mathrm{orb}=3.9 \, \text{hr}$. J1622 was discovered as a radio pulsar with a spin period of $3.86 \, \text{ms}$ \citep{sanpa2016searching} and a spin-down luminosity of $L_\mathrm{sd}=0.9\times10^{34} \, \text{erg/s}$ \citep{strader2019optical}. Based on the optical radial velocity curve of J1622, \cite{strader2019optical} found a low companion mass of $0.10$-$0.14 \, \text{M}_{\sun}$ and an inclination angle $i>64^{\circ}$ for a neutron star mass $1.5$-$2.0 \, \text{M}_{\sun}$. The inclination of J1622 was later constrained to be $i<83.4^{\circ}$ due to the absence of eclipses in its gamma-ray emission \citep{clark2023neutron}. In Section \ref{sec:2}, we describe the procedure used to perform data reduction and aperture photometry to extract the light curves of J1622. In Section \ref{sec:3}, we present and quantify an ellipsoidal flux modulation in the optical counterpart of J1622, indicating a lack of irradiation of the companion star, and compare it with other RBs. In Section \ref{sec:4}, we interpret our results in terms of a new parameter $f_\mathrm{sd}$, which can quantify the effect of the pulsar wind irradiation on the companion star.
\newpage
\section{Observations and data analysis}
\label{sec:2}
%\subsection{Instruments and data reduction}
%\label{sec:instranddatared} % used for referring to this section from elsewhere
We observed J1622 with the ALFOSC camera mounted on the 2.56-m Nordic Optical Telescope (NOT), located at the Spanish "Roque de los Muchachos" Observatory. The 2-min long exposures were acquired during 2022 April 21 in a $6.4 \, \arcmin\times6.4 \, \arcmin$ field-of-view, alternating the SDSS \textit{g'}, \textit{r'} and \textit{i'} filters for four consecutive hours. We also used a $2\times2$ binning for the CCD to reduce the readout times down to $8.1 \, \text{s}$ per exposure.
%The sequence of the procedure was: averaging all bias images into one final "master" bias, subtracting the master bias from all flat field and target images, combining all flat images into one master flat separately for each filter and finally dividing all images by the corresponding master flat.
%\subsection{Aperture photometry and comparison stars}
%\label{sec:aperturephot}
\begin{figure}
    \centering
    \includegraphics[width=\columnwidth]{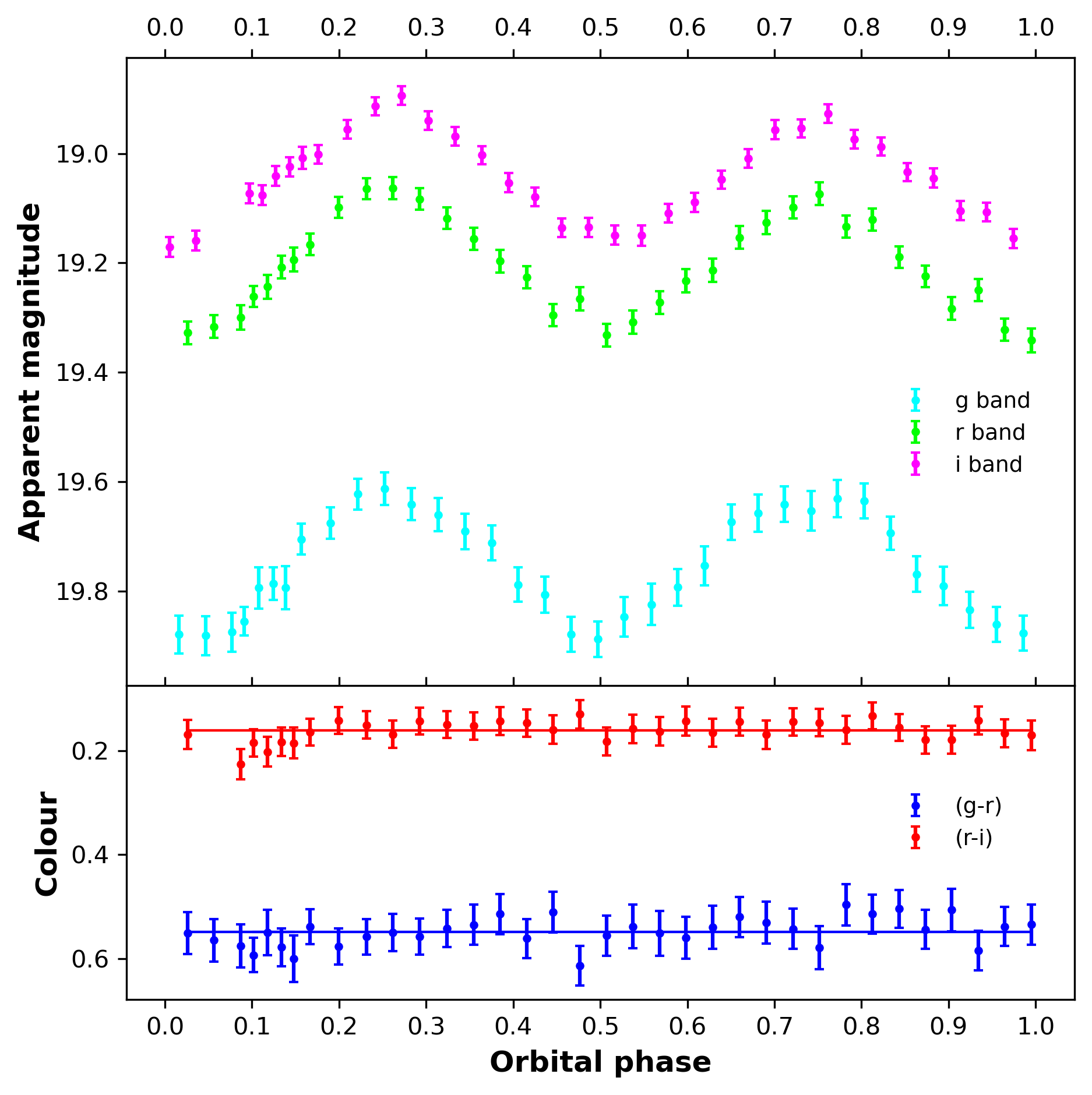}
    \caption{\textit{Top panel}: light curves of J1622 in the \textit{g'}, \textit{r'} and \textit{i'} optical bands folded at the orbital period $P_\mathrm{orb}=0.1617006798 \, \text{d}$ and reference epoch $T_\mathrm{0}=59691.05286 \, \text{MJD}$. \textit{Bottom panel}: observed colours (\textit{g'} - \textit{r'}) and (\textit{r'} - \textit{i'}), with solid lines showing a constant fitted to the data.}
    \label{fig:lcgri}
\end{figure}

After having reduced our images using the software package \textsc{IRAF}\footnote{\url{https://iraf-community.github.io/}}, we employed \textsc{ULTRACAM}\footnote{\url{https://cygnus.astro.warwick.ac.uk/phsaap/software/ultracam/html/index.html}} routines to perform differential aperture photometry of J1622 with three different sets of stable comparison stars (one for each optical filter \textit{g'}, \textit{r'} and \textit{i'}), setting the aperture radius to 1.5 times the seeing. We applied an ensemble photometry technique to improve the S/N ratio on the apparent magnitudes of our target \citep{honeycutt1992ccd}. Since the net error on the target magnitude is a combination of both the error in the target counts measurement and the error in the comparison counts measurement, this method consists in summing together the counts measured from $N$ stars to reduce the error contribution from this "master" comparison by a factor $1/\sqrt{N}$. The comparison stars, all showing small variability with rms amplitudes $\lesssim 0.03 \, \text{mag}$, were selected using the \textsc{astrosource}\footnote{\url{https://github.com/zemogle/astrosource}} package \citep{fitzgerald2021astrosource}, and combined to three master comparison stars that show even lower rms amplitudes $\simeq 0.01 \, \text{mag}$.
\section{Results}
\label{sec:3}
\subsection{Optical light curves of the redback PSR~J1622-0315}
\label{sec:optlcJ1622}

We show in Figure~\ref{fig:lcgri} the phase-folded \textit{g'}, \textit{r'} and \textit{i'}-band light curves as well as (\textit{g'} - \textit{r'}) and (\textit{r'} - \textit{i'}) colours of J1622. After converting the mid-exposure times of the optical images from UTC to TDB, we folded the light curves using the orbital period $P_\mathrm{orb}=0.1617006798(6) \, \text{d}$ \citep{sanpa2016searching} and as reference epoch the time of passage to the inferior conjunction of the companion $T_\mathrm{0}=59691.05286(1) \, \text{MJD}$\footnote{The time $T_\mathrm{0}$ was obtained propagating the corresponding parameter from the observing epoch of \cite{sanpa2016searching} to the epoch of our observations.}. Propagating the uncertainties on $P_\mathrm{orb}$ and $T_\mathrm{0}$ we get negligible errors of $\simeq10^{-4}$ on the orbital phases. As can be clearly seen from Figure~\ref{fig:lcgri}, we identify the same periodic modulation in all three bands, showing minima of the optical flux around the orbital phases $\phi=0$ and $\phi=0.5$ and maxima around $\phi=0.25$ and $\phi=0.75$, with a same peak-to-peak amplitude of $\simeq0.3 \, \text{mag}$ in \textit{g'}, \textit{r'} and \textit{i'} filters. 
%We also notice a phase asymmetry in the \textit{i'} light curve, with the first of the two maxima occurring at phase $\phi=0.27$ instead of $\phi=0.25$ like we observe in \textit{g'} and \textit{r'}.

The observed flux modulation reveals a lack of irradiation of the companion's inner face by the pulsar wind in this system. In an irradiated system we would expect to observe the minimum and maximum flux at the orbital phases $\phi=0$ and $\phi=0.5$ respectively, corresponding to inferior and superior conjunction when the companion star is showing its coldest and hottest side \citep{breton2013discovery}. In this case, the observed light curves can be attributed instead to the ellipsoidal shape of the companion. We see the two flux maxima at ascending and descending nodes (phases $\phi=0.25$ and $\phi=0.75$ respectively) due to the gravitationally distorted companion star.

We fitted the colours $(\textit{g}'-\textit{r}')$ and $(\textit{r}'-\textit{i}')$ observed along the orbit with a constant (as shown in the bottom panel of Figure~\ref{fig:lcgri}), finding for the two fits $\chi^{2}/\text{d.o.f.}=17.9/34$ and $18.1/33$, respectively\footnote{We considered only the statistical errors on the colors, excluding the uncertainties given by the magnitude calibration of the comparison stars.}. Both chi-square values support the hypothesis that the colours remain constant along the orbit. After dereddening J1622 apparent magnitudes using the colour excess value $E(g-r)=(0.23\pm0.02) \, \mathrm{mag}$ computed from the 3D dust map of \cite{green20193d}, we estimate the intrinsic colours along the orbit as $(\textit{g}'-\textit{r}')=0.2\,$--$\,0.5 \, \text{mag}$ and $(\textit{r}'-\textit{i}')=-0.1\,$--$\,0.1 \, \text{mag}$. According to Tab. 5 in \cite{pecaut2013intrinsic}, these colour intervals are compatible with a companion of spectral type $\text{F4V} \, $--$ \, \text{G1V}$, corresponding to an effective temperature $T_{2}=(6265\pm405)$~K, where we assumed that the optical emission is dominated by the companion star. %Then we also measured $(\textit{g}'-\textit{r}')=0.35(4) \, \text{mag}$ and $(\textit{r}'-\textit{i}')=0.03(3) \, \text{mag}$ at phase $\phi=0.5$, while  we find $(\textit{g}'-\textit{r}')=0.34(4) \, \text{mag}$ and $(\textit{r}'-\textit{i}')=0.02(3) \, \text{mag}$ correspondingly to the inferior conjunction. From these colour values, adopting the same procedure used above, we infer the temperature $T_\mathrm{h}\simeq 6280 \, \text{K}$ of the companion inner side facing the pulsar, while we find $T_\mathrm{b}\simeq 5990 \, \text{K}$ for the outer side.
Therefore, the difference between the temperature of the companion inner side $T_\mathrm{h}$ and the temperature of the outer side $T_\mathrm{b}$ is $\Delta T\lesssim500$~K in J1622, that is smaller than typically measured for irradiated RBs $\Delta T\gtrsim10^{3}$~K \citep{breton2013discovery}. %Another relevant parameter one can compute is the irradiation temperature, defined as $T_\mathrm{irr}^{4}=T_\mathrm{h}^{4}-T_\mathrm{b}^{4}$, which is directly dependent on the fraction of pulsar wind flux hitting the companion surface. We estimate $T_\mathrm{irr}\simeq4046 \, \text{K}$ for J1622, value lower respect $T_\mathrm{irr}>5000 \, \text{K}$ typically measured from irradiated redbacks \citep{breton2013discovery}.
We conclude that the temperature change between inner and outer face of the companion is too faint to be detected with the sensitivity of our observations.
%For this system \cite{clark2023neutron} recently estimated an upper limit on the inclination of the orbital plane $i<83.4^{\circ}$, based on the absence of observed gamma-ray eclipses. However, since this constraint turns out to be quite loose we decided to shrink the interval   
\newpage
\subsection{Comparison with other redbacks}
\label{sec:J2215}
\begin{figure}
    \centering
    \includegraphics[width=\columnwidth]{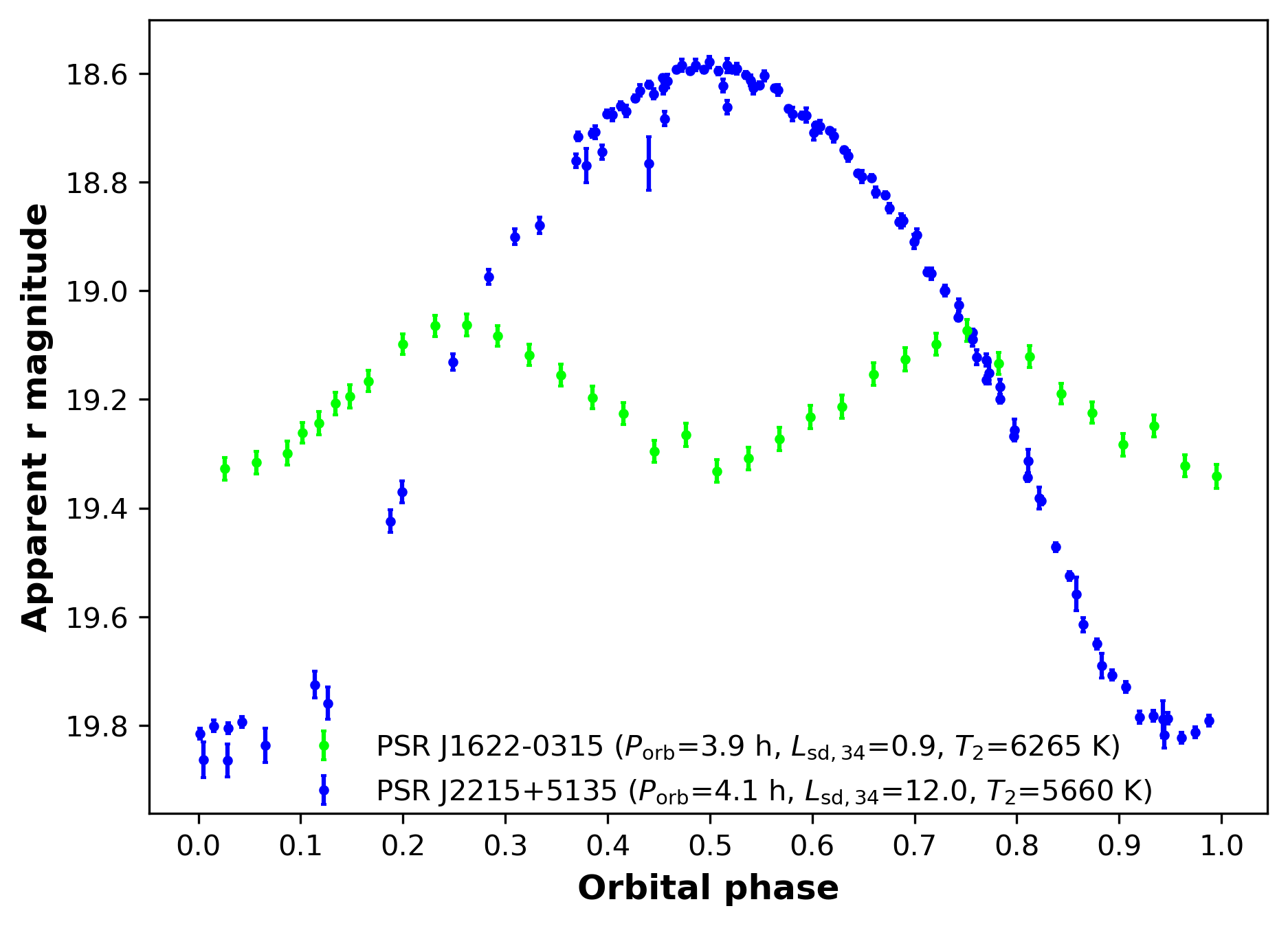}
    \caption{\textit{In green}: light curve in the \textit{r'} band of PSR~J1622-0315 obtained from our analysis showing ellipsoidal modulation. \textit{In blue}: light curve in the \textit{r'} band of PSR~J2215+5135 obtained by \protect\cite{linares2018peering} showing an irradiation-dominated modulation. The spin-down luminosities for the two systems are reported here in units of $10^{34} \, \text{erg/s}$.}
    \label{fig:compareJ2215}
\end{figure}

We searched in the literature public optical light curves of all known RBs and BWs (about 60 systems at the time of writing). We find that the same type of modulation showing absence of irradiation has been observed in about $50\%$ of the known RB population. For instance, PSR~J1431-4715 \citep[$P_\mathrm{orb}=10.8 \, \text{h}$;][]{strader2019optical}, PSR~J1723-2837 \citep[$P_\mathrm{orb}=14.8 \, \text{h}$;][]{van2016psr}, PSR~J2129-0429 \citep[$P_\mathrm{orb}=15.2 \, \text{h}$;][]{al2018x} and  PSR~J0212+5321 \citep[$P_\mathrm{orb}=20.9 \, \text{h}$;][]{linares2017millisecond}, all show optical light curves dominated by ellipsoidal modulation (qualitatively similar to J1622). However, in all these systems the wider orbits likely cause a significant decrease of the fraction of the pulsar's spin-down luminosity impinging on the surface of the companion. Therefore, it is surprising that we observe the same in J1622 despite its short orbital period, which makes this system the most compact known RB without an irradiation-dominated optical light curve.

We compared the light curve of J1622 with the RB PSR~J2215+5135 (henceforth J2215), both shown in Figure~\ref{fig:compareJ2215}. Although the orbital periods are similar, %of J2215 $P_\mathrm{orb}=4.1 \, \text{h}$ is really close to the $P_\mathrm{orb}=3.9 \, \text{h}$ of J1622,
the two systems show drastically different behaviour. Indeed, as shown in Figure~\ref{fig:compareJ2215}, the optical light curve of J2215 has only one flux maximum per orbit occurring at $\phi=0.5$, with an \textit{r'}-band peak-to-peak amplitude of $\simeq1 \, \text{mag}$ \citep{linares2018peering}, as opposed to the two maxima at $\phi=0.25/0.75$ and the much lower $\simeq0.3 \, \text{mag}$ amplitude seen in J1622. In J2215, the companion star is heavily irradiated by the pulsar wind \citep{breton2013discovery,linares2018peering}, opposite to what is observed in J1622. %\cite{linares2018peering} estimated for J2215 a temperature of the companion irradiated side $T_\mathrm{h}=8080 \, \text{K}$ much higher respect the base (dark side) temperature $T_\mathrm{b}=5660 \, \text{K}$, while its spin-down luminosity is $L_\mathrm{sd}=1.2\times10^{35} \, \text{erg/s}$ \citep{strader2019optical}. The base temperature and spin-down luminosity turn out to be quite different respect the corresponding values measured for J1622, respectively $T_{2}=6265 \, \text{K}$ (from our analysis) and $L_\mathrm{sd}=0.9\times10^{34} \, \text{erg/s}$ \citep{strader2019optical}.
The differences between these two systems are the temperature difference along the orbit \citep[$\Delta T\simeq2400$~K in J2215;][]{linares2018peering} and the spin-down luminosity ($L_\mathrm{sd,J2215}=1.2\times10^{35} \, \text{erg/s}$; \citealt{strader2019optical}, $L_\mathrm{sd,J1622}=0.9\times10^{34} \, \text{erg/s}$; \citealt{strader2019optical}).

This might point towards the companion intrinsic brightness and the intensity of the pulsar wind contributing to the presence or absence of companion star irradiation in spiders in addition to the orbital period.

\section{Quantifying irradiation in spiders}
\label{sec:4}
%%%Full population: intro

\begin{figure}
%\centering
  \begin{center}
  %\resizebox{1.0\columnwidth}{!}{\rotatebox{-90}{\includegraphics[]{Teff-Mcmin.eps}}}
  \resizebox{1.0\columnwidth}{!}{{\includegraphics[]{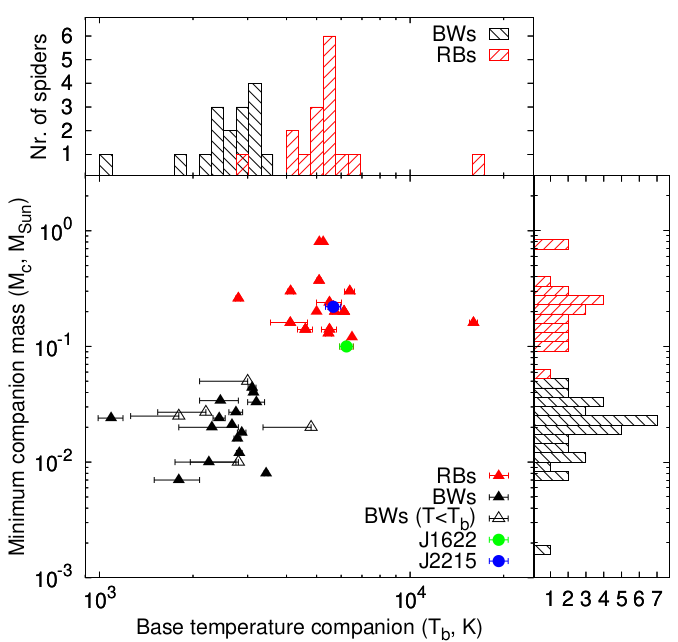}}}
  \caption{
Minimum companion mass from radio timing ($M_\mathrm{c}$) vs. ``base''
effective temperature of the companion ($T_\mathrm{b}$, from the
dark/outer/night side when irradiation is present). Top and side
panels show the corresponding histograms. BWs and RBs are shown as black and red triangles, respectively (open triangles show $T_\mathrm{b}$ upper limits). The green and the blue circles correspond to J1622 and J2215, respectively. 
} %
    \label{fig:Teff}
%\epsscale{1.0}
 \end{center}
\end{figure}

We now look at the full population of CBMSPs, in order to interpret
our results for J1622 and discuss their implications.
From a literature search of all currently known CBMSPs in the Galactic
field\footnote{We do not study here the globular cluster CBMSP
  population, since their optical properties are not so well
  characterised.}, we compiled the base effective temperature of the cold (not irradiated) side of the companion star, $T_\mathrm{b}$, as well as
the number of maxima in the optical light curve per orbital cycle,
$N_\mathrm{max}$ (Table~\ref{table:spiders}).
%
%%%Base temperature: redbacks are hotter
%Figure 3: TeffMcMin.ps
Figure~\ref{fig:Teff} shows that the intrinsic temperatures of the
companion stars in RBs are systematically higher than those of BW companions.
RB companions have typically $T_\mathrm{b} \simeq$4000--6000~K, similar to
isolated low-mass main sequence stars with spectral types around G
(from late K to early F).
BW companions, on the other hand, are ten times less massive and
about two times colder than RB companions, with typical base temperatures of $T_\mathrm{b}
\simeq$1000--3000~K \citep[similar to brown dwarfs with spectral types T-L-M,][]{basri2000observations}.

As explained in Section \ref{sec:3}, irradiation-dominated optical light curves
have $N_\mathrm{max} = 1$, while those where ellipsoidal modulation
dominates have $N_\mathrm{max} = 2$.
Strong irradiation and ablation characterised the first discovered BWs \citep{fruchter1988millisecond}, and indeed we see
that the optical light curves of all currently known BWs (15
confirmed BWs and 3 BW candidates) are always dominated by
irradiation ($N_\mathrm{max} = 1$).
\begin{table}
%\small
\scriptsize
%\tiny
\caption{Main properties of spiders used in this work: spin-down
  luminosity, orbital period ($P_\mathrm{orb}$), base temperature ($T_\mathrm{b}$) and pulsar-to-companion flux ratio ($f_\mathrm{sd}$; see text for definition). The last column shows the number of maxima in the optical light curve per orbital
  cycle ($N_\mathrm{max}$), when available. The population is split into RBs (top part) and BWs (bottom part).}
\begin{minipage}{0.5\textwidth}
%\centering
\setlength{\tabcolsep}{7.5pt} % Default value: 6pt
\begin{tabular}{lcccccc}
\hline\hline
\footnotetext{
  References $T_\mathrm{b}$:
  (1) \citet{shahbaz2022peculiar};
  (2) \citet{yap2019face};
  (3) \citet{de2015multiwavelength};
  (*) This work;
  (4) \citet{li2014optical};
  (5) \citet{crawford2013psr};
  (6) \citet{kaplan2013metal};
  (7) \citet{clark2021einstein};
  (8) \citet{bellm2016properties}
  (9) \citet{linares2018peering};
  (10) \citet{romani2011orbit};
  (11) \citet{draghis2019multiband};
  (12) \citet{dhillon2022multicolour};
  (13) \citet{mata2023black};
  (14) \citet{romani2012psr};
  (15) \citet{kandel2022optical};
  (16) \citet{zharikov2019optical};
  (17) \citet{breton2013discovery}.
  Methods $T_\mathrm{b}$:
  base effective temperature of the companion star, as inferred
  from: i-optical light curve fits / ii-optical colors / iii-optical
  spectra: optimal subtraction (or line ID) / iv-optical spectra:
  modelling / v: broadband SED fit.
}    
Name & $L_\mathrm{sd}^\mathrm{a}$ & $P_\mathrm{orb}$ & $T_\mathrm{b}$ & REF & $f_\mathrm{sd}$ & $N_\mathrm{max}$\\
PSR & ($10^{34}$~erg~s$^{-1}$) & (h) & (K) & (met.) &  & \\
\hline
J1023+0038 & 4.8$\pm$0.5 & 4.8 &	5724$\pm$100 & 1-iv	& 3.87 & 1 \\
J1048+2339 & 1.2$\pm$0.1 & 6.0 &	4123$\pm$80 &  2-i	& 2.59 & 1.5\\
J1227-4853 & 8.3$\pm$1.8 & 6.9 &	5500$\pm$300 & 3-i	& 4.94 & 1 \\
J1622-0315 & 0.9$\pm$0.1 & 3.9 &	6265$\pm$405 & *-ii	& 0.69 & 2 \\
J1628-3205 & 1.4 & 5.0 &	4115$\pm$555 & 4-i & 4.07 & 2 \\
J1723-2837 & 2.5$\pm$0.5 & 14.8 &	5500$\pm$500 & 5-iii/v	& 0.52 & 2 \\
J1816+4510 & 7.3$\pm$0.7 & 8.7 &	16000$\pm$500 & 6-iv	& 0.044 & ? \\
J2039-5618 & 2.5$\pm$0.5 & 5.4 &	5460$\pm$140 & 7-i	& 2.11 & 2 \\
J2129-0429 & 4.8$\pm$0.5 & 15.2 &	5094$\pm$90 & 8-i	& 1.26 & 2 \\
J2215+5135 & 12.0$\pm$0.9 & 4.1 &	5660$\pm$320 & 9-iii	& 12.4 & 1 \\
J2339-0533 & 2.5$\pm$0.4 & 4.6 &	2800$\pm$50 & 10-i	& 36.6 & 1 \\
\hline
B1957+20 & 9.3$\pm$0.3 & 9.2 &	2670$\pm$30 & 11-i	& 70.9 & 1 \\
J2051-0827 & 0.46$\pm$0.05 & 2.4 &	2750$\pm$140 & 12-i	& 18.7 & 1 \\
J0023+0923 & 1.43$\pm$0.04 & 3.3 &	2780$\pm$70 & 13-i	& 36.5 & 1 \\
J0251+2606 & 1.1$\pm$0.3 & 4.85 &	1090$\pm$100 & 13-i	& 709 & 1 \\
J0636+5128 & 0.55$\pm$0.01 & 1.6 &	1800$\pm$300 & 13-i	& 210 & 1 \\
J1301+0833 & 6.7 & 6.5 &	2430$\pm$110 & 11-i	& 117 & 1 \\
J1311-3430 & 5.053$\pm$0.004 & 1.56 &	3440$\pm$50 & 14-?	& 150 & 1 \\
J1544+4937 & 1.42$\pm$0.02 & 2.8 &	2870$\pm$90 & 13-i	& 39.6 & 1 \\
J1641+8049 & 4.28$\pm$0.04 & 2.2 &	3130$\pm$80 & 13-i	& 115 & 1 \\
J1653-0159 & 0.45$\pm$0.04 & 1.25 &	2250$\pm$500 & 15-i	& 97.8 & 1 \\
J1810+1744 & 4.18 & 3.6 &	3100$\pm$90 & 15-i	& 60.8 & 1 \\
J2052+1219 & 1.7$\pm$0.2 & 2.75 &	3200$\pm$200 & 16-i	& 31.3 & 1 \\
J2241-5236 & 1.76$\pm$0.04 & 3.5 &	2820$\pm$60 & 11-i	& 39.2 & 1 \\
J2256-1024 & 3.7$\pm$0.1 & 5.1 &	2450$\pm$350 & 17-i	& 86.8 & 1 \\
\hline
  \end{tabular}
  \footnotesize{$^\mathrm{a}$ Spin-down luminosities values, taken from \cite{strader2019optical} for RBs and \cite{mata2023black} for BWs, include the Shklovskii effect correction where the proper motion of the system is known.}
%  \begin{tablenotes}
%    References:
%    1 \citet{Shahbaz22};
%    2 \citet{Yap19};
%    3 \citet{deMartino15}
%  \end{tablenotes}
\end{minipage}
\label{table:spiders}
\end{table}

\begin{figure*}
%\centering
  \begin{center}
  %\resizebox{2.0\columnwidth}{!}{\rotatebox{-90}{\includegraphics[]{OLC.eps}}}
  \resizebox{2.0\columnwidth}{!}{{\includegraphics[]{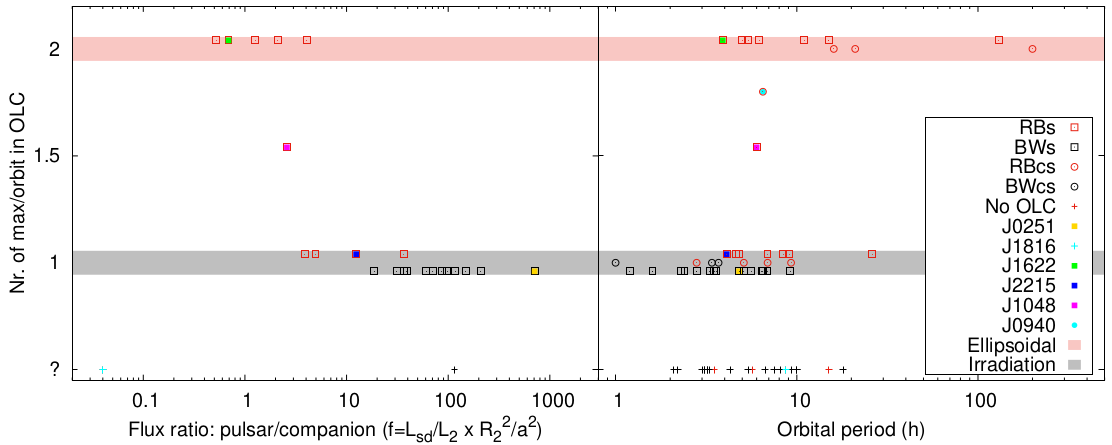}}}
  \caption{
Number of maxima in the optical light curve per orbital cycle
($N_\mathrm{max}$) plotted as a function of pulsar/companion flux ratio ($f_\mathrm{sd}$, left; see text for definition) and orbital period ($P_\mathrm{orb}$, right panel).
RBs are shown in red symbols, BWs in black symbols. Open squares and
circles show confirmed and candidate systems, respectively, with small
arbitrary shifts from $N_\mathrm{max}=1,2$ for display purposes.
Five RBs and one BW discussed in the text are highlighted with different
colors, as indicated.
Systems with unknown $N_\mathrm{max}$ (due to lack of optical
counterpart or poor data quality) are shown with red/black plus signs
along the bottom axis (and marked with ``?'').
} %
    \label{fig:OLC}
%\epsscale{1.0}
 \end{center}
\end{figure*}

We find, however, that about half of the currently known RBs have
$N_\mathrm{max} = 1$ (7 confirmed RBs, 4 RB candidates), while the
other half show little or no signs of irradiation, with light curves
dominated by ellipsoidal modulation and $N_\mathrm{max} = 2$ (8
confirmed RBs, 2 candidates).
This is shown in Figure~\ref{fig:OLC} (right panel), where we show $N_\mathrm{max}$
vs. $P_\mathrm{orb}$.
%

%Table 1 IF WE HAVE SPACE?

%%%What determines the importance of irradiation?

We have established in Section \ref{sec:J2215} that orbital separation alone cannot
explain the presence or absence of irradiation inferred from the
optical light curves.
Here, we argue that the combination of three different factors
determines the importance of irradiation in spiders, namely:

{\it i) pulsar.} The irradiating source is powered by the loss of
kinetic energy of rotation of the neutron star, and thus its intensity
is set by $L_\mathrm{sd}$. X-ray and gamma-ray luminosities, as well as those
in electron-positron pairs from the magnetosphere \citep{harding2011pulsar}
and intra-binary shock \citep{linares2021cosmic}, all scale positively with $L_\mathrm{sd}$.

{\it ii) Orbit.} For a given $L_\mathrm{sd}$, the spin-down energy flux
impinging on the companion star is inversely proportional to the
square of the orbital separation between pulsar (irradiating
source) and companion ($a$). CBMSPs with short $P_\mathrm{orb}$
are thus expected to show strong irradiation (Section \ref{sec:J2215}).

{\it iii) Companion.} The intrinsic stellar flux from the companion will also determine the importance of
irradiation. A hotter companion will be more difficult to irradiate
than a cooler star.

Combining these three factors, we define the pulsar spin-down to
companion flux ratio at the companion as
%
%f$_{sd}$
%
\begin{equation}
\label{eq:fsd}
f_\mathrm{sd} \equiv \frac{L_\mathrm{sd}}{L_2} \frac{R_2^2}{a^2} ,
\end{equation}
where $R_{2}$ is the companion radius and $L_{2}$ its intrinsic luminosity
(before/without irradiation).
This dimensionless ratio provides a simple way to quantify the
importance of irradiation, as previously discussed for dwarf novae
\citep{wade1988radial} and low-mass X-ray binaries \citep{phillips1999outburst}. 
\citet{linares2018peering} applied a similar estimate to J2215, but using the
irradiating luminosity estimated from optical light/radial-velocity
curve fits ($L_\mathrm{irr}$), instead of the more fundamental and directly measurable
$L_\mathrm{sd}$.

Assuming a spherical companion with an intrinsic (unirradiated)
temperature $T_\mathrm{b}$, we can approximate
\begin{equation}
\label{eq:l2}
L_2 \simeq 4 \pi R_2^2 \sigma T_\mathrm{b}^4.
\end{equation}

Using Kepler's third law we can relate $a$ with the total mass in the
binary ($M=M_1 + M_2$) and $P_\mathrm{orb}$: $a^2 \propto M^{2/3}
P_\mathrm{orb}^{4/3}$.
%
%\begin{equation}
%\label{eq:kepler}
%a^2 \propto M^{2/3} P_\mathrm{orb}^{4/3},
%\end{equation}
%
Then Equation~\ref{eq:fsd} turns into:
\begin{equation}
  \begin{aligned}
%  \begin{split}
\label{eq:fsdcalc}
& f_\mathrm{sd} \simeq 
\frac{1}{4\pi\sigma} \left(\frac{4\pi^2}{G}\right)^{2/3} L_\mathrm{sd} T_\mathrm{b}^{-4} M^{-2/3} P_\mathrm{orb}^{-4/3} =\\
& 1.1\times10^4 \left[\frac{L_\mathrm{sd}}{10^{34}~erg~s^{-1}}\right] \left[\frac{T_\mathrm{b}}{10^{3}~K}\right]^{-4} \left[\frac{M}{M_\odot}\right]^{-\frac{2}{3}} \left[\frac{P_\mathrm{orb}}{1~h}\right]^{-\frac{4}{3}}.
 \end{aligned}
% \end{split}
\end{equation}
Since $M$ can vary by a factor $\lesssim$2 and has the weakest effect on
$f_\mathrm{sd}$, we can assume an intermediate value $M \simeq 1.8
M_\odot$ and reach a simpler analytical approximation:
\begin{equation}
%  \begin{aligned}
%  \begin{split}
\label{eq:fsdest}
f_\mathrm{sd} \simeq 7700 \left[\frac{L_\mathrm{sd}}{10^{34}~erg~s^{-1}}\right] \left[\frac{T_\mathrm{b}}{10^{3}~K}\right]^{-4} \left[\frac{P_\mathrm{orb}}{1~h}\right]^{-\frac{4}{3}}
% \end{aligned}
% \end{split}
\end{equation}
(which agrees with Equation~\ref{eq:fsdcalc} within about 10\%).
In this derivation we have ignored the effects of the companion size
(implicitly assuming $a >> R_2$) but its inner/day side will receive a
higher irradiating flux than its outer/night side.
To quantify this correction, assuming that the companion star fills
its Roche lobe, we evaluate Equation~\ref{eq:fsd} at $a-R_2$ instead of $a$,
and find the same trends with a 35\% increase in $f_\mathrm{sd}$. Although anisotropic pulsar winds may arise depending on the angle between the pulsar spin and the magnetic axes \citep{spitkovsky2006time,philippov2015ab}, in Equation~\ref{eq:fsd} we assume for simplicity isotropic emission of the spin-down luminosity.

%%%Our new parameter explains presence/absence of irradiation.

Armed with this new parameter, we calculate $f_\mathrm{sd}$ for the
full spider population, using Equation~\ref{eq:fsdcalc} and the best
measurements of $P_\mathrm{orb}$, $L_\mathrm{sd}$, $M$ and
$T_\mathrm{b}$ available in the literature.
The results are shown in Figure~\ref{fig:OLC} (left panel), where we show
$N_\mathrm{max}$ vs. $f_\mathrm{sd}$.
The range is wide, more than 4 orders of magnitude in $f_\mathrm{sd}$ (as we can see in Table~\ref{table:spiders}),
showing the widely different irradiation conditions at work in
different spider companions.
We find that $f_\mathrm{sd}$ explains the dichotomy between irradiated and non-irradiated RBs: low $f_\mathrm{sd}$ results into ellipsoidal light curves without signs of irradiation (and $N_\mathrm{max}=2$), whereas high $f_\mathrm{sd}$ results into irradiation-dominated light curves ($N_\mathrm{max}=1$).
This transition occurs at $f_\mathrm{sd} \simeq 2$--$4$, which explains the fact that all known BWs are dominated by irradiation: they all have $f_\mathrm{sd} > 18$.
The pulsar-to-companion flux ratios measured for J1622 and J2215, $f_\mathrm{sd}=0.7$ and $f_\mathrm{sd}=12$ respectively, are therefore able to explain the different light curves we observe for the two systems (see Section \ref{sec:J2215}).
\begin{figure*}
%\centering
  \begin{center}
  %\resizebox{2.0\columnwidth}{!}{\rotatebox{-90}{\includegraphics[]{Edot-Porb-f.eps}}}
  \resizebox{2.0\columnwidth}{!}{{\includegraphics[]{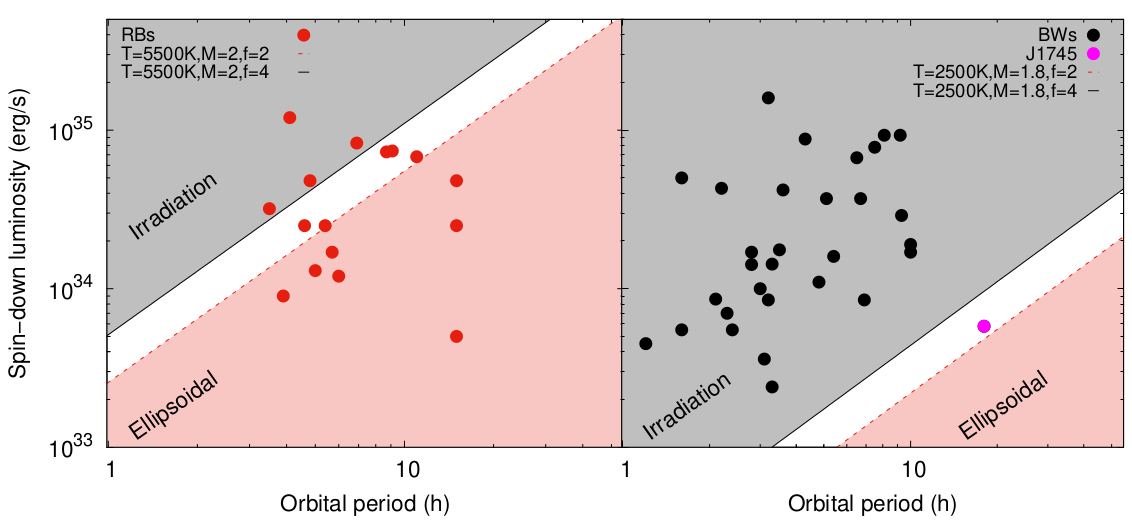}}}
  \caption{
Spin-down luminosity ($L_\mathrm{sd}$) vs. orbital period
($P_\mathrm{orb}$) for RBs (left) and BWs (right).
Black solid and red dashed lines show the $L_\mathrm{sd} \propto
P_\mathrm{orb}^{4/3}$ relations for $f_\mathrm{sd}=2$ and
$f_\mathrm{sd}=4$, respectively, using fixed values of total mass ($M$) and
base temperature ($T_\mathrm{b}$).
We chose representative values of hotter and more massive RB
companions ($T_\mathrm{b}=5500$~K, $M = 2 M_\odot$) and colder, less
massive BWs companions ($T_\mathrm{b}=2500$~K, $M = 1.8
M_\odot$).
In either case, we highlight regions dominated by irradiation
($f_\mathrm{sd} > 4$, gray shaded area) and those where ellipsoidal
modulation prevails ($f_\mathrm{sd} < 2$, red shaded area).
} %
    \label{fig:Edot}
%\epsscale{1.0}
 \end{center}
\end{figure*}

%Intermediate systems: J1048, J0940
%\begin{figure}
%    \includegraphics[width=\columnwidth]{etavsfsd.png}
%    \caption{From the top to the bottom panel are shown respectively $T_\mathrm{irr}$, $L_\mathrm{irr}$ and $L_\mathrm{irr}/L_\mathrm{sd}$ as a function of $f_\mathrm{sd}$ for irradiated systems (redbacks in red and black widows in black), with the corresponding Spearman's correlation coefficients and p-values. Redback systems are highlighted with their orbital inclinations $i$ taken from \cite{strader2019optical} and references therein.}
%    \label{fig:etavsfsd}
%\end{figure}
%\begin{figure}
%    \includegraphics[width=\columnwidth]{etavsincl.png}
%    \caption{From the top to the bottom panel are shown respectively $T_\mathrm{irr}$, $L_\mathrm{irr}$ and $L_\mathrm{irr}/L_\mathrm{sd}$ as a function of the inclination $i$ for irradiated redbacks, with the corresponding Spearman's correlation coefficients and p-values.}
%    \label{fig:etavsincl}
%\end{figure}
Two peculiar RBs are interesting to discuss in this context: PSR~J1048+2339 and 4FGL~J0940.3-7610. PSR~J1048+2339 ($P_\mathrm{orb}$=6~h, $L_\mathrm{sd}$=1.2$\times$10$^{34}$~erg~s$^{-1}$) switched from an ellipsoidal to an irradiated light curve on timescales of days-weeks \citep[from $N_\mathrm{max}$=2 to $N_\mathrm{max}$=1,][]{yap2019face}.
The RB candidate 4FGL~J0940.3-7610 ($P_\mathrm{orb}$=6.5~h) showed two maxima, but significant brightening at superior
conjunction was observed compared with the absolute minimum at inferior conjunction
of the companion \citep[by about 0.3 mag in g';][see their Figure~3]{swihart2021discovery}.
This suggests that its inner face is mildly irradiated.
For these reasons, we arbitrarily assign values of $N_\mathrm{max}$=1.5 and
1.75 to PSR~J1048+2339 and 4FGL~J0940.3-7610, respectively, and we
suggest that they represent a regime of intermediate irradiation.
Indeed, we find $f_\mathrm{sd}=2.6$ for PSR~J1048+2339, consistent
with being at the transition between ellipsoidal and
irradiation-dominated light curves.

In Figure~\ref{fig:OLC} (left panel) we also see two extreme cases: the RB
PSR~J1816+4510, with the lowest $f_\mathrm{sd}=0.04$, should show no
signs of irradiation %\citep[due to its very high $T_\mathrm{b}$][the optical light curves are not conclusive]{kaplan2012discovery}.
(as confirmed from its ellipsoidal optical light curves; Koljonen \& Linares 2023, in prep.).
On the other hand, the BW PSR~J0251+2606 has the highest
$f_\mathrm{sd}=709$ and indeed shows an extremely large magnitude
difference between the day and night side \citep[at least
  3-4~mags;][]{draghis2019multiband,mata2023black}.

Assuming a typical $M$ and $T_\mathrm{b}$, we can now use
Equation~\ref{eq:fsdest} to predict the importance of irradiation for
newly discovered spiders, measuring only $P_\mathrm{orb}$ and
$L_\mathrm{sd}$.
This is shown in Figure~\ref{fig:Edot}, where we plot the regions dominated by
irradiation ($f_\mathrm{sd} > 4$, gray shaded area) and those where
ellipsoidal modulation prevails ($f_\mathrm{sd} < 2$, red shaded
area), both for RBs (left) and BWs (right). In this context it is worth mentioning the BW PSR~J1745+1017 ($P_\mathrm{orb}$=18~h, $L_\mathrm{sd}$=5.8$\times$10$^{33}$~erg~s$^{-1}$), discovered as a radio pulsar by \cite{barr2013pulsar} but without any optical counterpart reported yet. As we can see from Figure~\ref{fig:Edot} (right panel), this system departs from all the other BWs and stands very close to the ellipsoidal (red) region. We therefore predict that it might be the first case of a non-irradiated BW.

\section{Conclusions}
\label{sec:5}
In this paper, we presented multi-band optical light curves of the RB PSR~J1622-0315, showing two flux maxima and no detectable temperature changes along the orbit, clearly indicating the lack of irradiation of the companion star by the pulsar wind despite the short orbital period. %This is completely different to what we see in the RB PSR~J2215+5135, having an orbital period close to the one of PSR~J1622-0315 ($P_\mathrm{orb}=4.1 \, \text{h}$ and $P_\mathrm{orb}=3.9 \, \text{h}$ respectively) yet showing a strongly irradiated optical light curve, with only one maximum per orbital cycle.
We attribute this behaviour to a low value of the pulsar spin-down to companion flux ratio $f_\mathrm{sd}=0.7$, which characterises the strength of the companion star irradiation.

We extend our study to the population of all known spiders, finding out that the presence or absence of irradiation in such systems is unambiguously determined by $f_\mathrm{sd}$, with a transition from ellipsoidal modulation to irradiated regime taking place at $f_\mathrm{sd} \simeq 2$--$4$. This new parameter depends on the orbital period $P_\mathrm{orb}$ of the system, the companion base temperature $T_\mathrm{b}$ and the pulsar spin-down luminosity $L_\mathrm{sd}$, and can be used to determine the expected light curve behaviour assuming isotropic pulsar wind emission. Future studies of the irradiation strength in a large sample of spiders may be able to probe the geometry of the pulsar wind.
%The simple and directly measurable parameter $f_\mathrm{sd}$ provided in this work is important to understand which optical light curve we expect for newly discovered spiders. Namely, measuring only $P_\mathrm{orb}$ and $L_\mathrm{sd}$ of the system, one can use Equation~\ref{eq:fsdest} to predict if the optical emission will be dominated by irradiation or ellipsoidal modulation, as Figure~\ref{fig:Edot} shows.

The pulsar-to-companion flux ratio represents also a useful tool to predict extreme cases of spiders, either with low ($f_\mathrm{sd}<1$) or high irradiation ($f_\mathrm{sd}>10$), which are the most suitable systems to obtain accurate measurements of the neutron star mass from the optical light curve modeling.

\section*{Acknowledgements}

This project has received funding from the European Research Council (ERC) under the European Union’s Horizon 2020 research and innovation programme (grant agreement No. 101002352). We thank S.S. Lindseth and A. Tidemann for performing the observations with NOT/ALFOSC in remote mode. We acknowledge the late T.R. Marsh for the use of ULTRACAM software. M. Linares thanks D. Russell for a discussion on irradiation and companion temperature.

\section*{Data availability}

The raw NOT images with bias and flats frames used for data reduction can be obtained by contacting M. Turchetta.

%%%%%%%%%%%%%%%%%%%%%%%%%%%%%%%%%%%%%%%%%%%%%%%%%%
%\section*{Data Availability}

%The inclusion of a Data Availability Statement is a requirement for articles published in MNRAS. Data Availability Statements provide a standardised format for readers to understand the availability of data underlying the research results described in the article. The statement may refer to original data generated in the course of the study or to third-party data analysed in the article. The statement should describe and provide means of access, where possible, by linking to the data or providing the required accession numbers for the relevant databases or DOIs.



\bibliographystyle{mnras}
\bibliography{example} % if your bibtex file is called example.bib

%\begin{thebibliography}{99}
%\end{thebibliography}

%%%%%%%%%%%%%%%%%%%%%%%%%%%%%%%%%%%%%%%%%%%%%%%%%%

%%%%%%%%%%%%%%%%% APPENDICES %%%%%%%%%%%%%%%%%%%%%

%\appendix



%%%%%%%%%%%%%%%%%%%%%%%%%%%%%%%%%%%%%%%%%%%%%%%%%%

% Don't change these lines
\bsp	% typesetting comment
\label{lastpage}
\end{document}